\documentclass[aps,twocolumn,groupedaddress]{revtex4}

\usepackage{graphicx }

\begin{document}
\bibliographystyle{apsrev}


\title{Evaporative cooling of cesium atoms in the
  gravito-optical surface trap}



\author{M.~Hammes}
\author{D.~Rychtarik}
\author{R.~Grimm}
\affiliation{Institut f\"ur Experimentalphysik, Universit\"at
Innsbruck, Technikerstra{\ss}e 25, 6020 Innsbruck, Austria}

\date{12 November 2000}

\begin{abstract}
\begin{center}
\end{center}
We report on cooling of an atomic cesium gas closely above an
evanescent-wave atom mirror. Our first evaporation experiments
show a temperature reduction from 10\,$\mu$K down to 300\,nK along
with a gain in phase-space density of almost two orders of
magnitude. In a series of measurements of heating and spin
depolarization an incoherent background of resonant photons in the
evanescent-wave diode laser light was found to be the limiting
factor at this stage.
\end{abstract}

\maketitle

\section{Introduction}
Optical dipole traps based on far-detuned laser light have become
very popular as versatile tools for experiments on ultracold
atomic gases in an almost dissipation-free environment
\cite{gri00}. In our experiments, we use the {\em gravito-optical
surface trap} (GOST) \cite{ovc97} to produce an ultracold sample
of cesium atoms on an evanescent-wave atom mirror with the
prospect to study a two-dimensional atomic gas at high phase-space
densities.

Atomic cesium is a candidate of particular interest because of its
resonant scattering properties. In the lowest Zeeman-substate, a
low-field Feshbach resonance \cite{vul99a} facilitates convenient
magnetic tuning of the $s$-wave scattering length from zero to
very large positive or negative values. For low-dimensional atomic
systems, a scattering length exceeding the extension of the
ground-state wavefunction in one or more dimensions constitutes an
intriguing system \cite{ols98,pet00}.

Here, after summarizing the basic properties of the trap
(Sec.\ref{secGOST}), we report our first experiments demonstrating
evaporative cooling of atoms in the GOST (Sec.\ref{secEVAP}). This
important step to attain a two-dimensional gas at high phase-space
densities is achieved by ramping down the optical trapping
potentials. The current limitations imposed by heating in the GOST
are investigated (Sec.\ref{secHEAT}), and the prospects of future
experiments are discussed (Sec.\ref{secOUTLOOK}).

\section{Gravito-optical surface trap}\label{secGOST}
\begin{figure}[tb]
\begin{center}
\includegraphics[width=6cm]{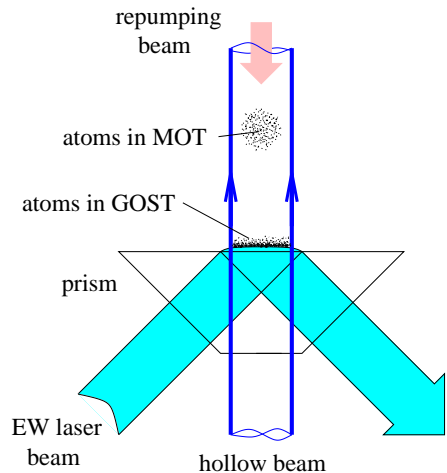}
\end{center}
\caption{Illustration of the gravito-optical surface trap.}
\label{gost} \vspace{3mm}
\end{figure}

A schematic overview of the geometry of the trap is given in figure
\ref{gost}. The GOST is an ``optical mug'', whose bottom consists of an
evanescent-wave (EW) atom mirror generated by total internal reflection
of a blue-detuned laser beam from the surface of a prism, while the
walls are formed by an intense hollow beam (HB) which passes vertically
through the prism surface.

The steep exponential decay  of the EW intensity along the
vertical direction and the sharp focussing of the hollow beam lead
to large intensity gradients and thus in combination with the blue
detuning of both light fields to a strong repulsive dipole force.
We exploit this fact to efficiently keep the atoms in the dark
inner region of the trap where the unwanted effect of heating
through scattering of trapping light photons is suppressed. In
addition to that, the concept also features a large trapping
volume which allows for a transfer of a large number of atoms into
the GOST. Due to the accurate focussing of the hollow laser beam
into a ring-shaped intensity profile \cite{man98}, the shape of
the potential is box-like along the horizontal directions whereas
the combination of gravity and the repulsive wall of the EW leads
to a wedge-shaped potential vertically.

The experimental constituents of the GOST are the EW diode laser
(SDL-5712-H1, distributed Bragg reflector), a titanium:sapphire laser
to create the hollow beam and an additional diode laser to provide the
light for the repumping beam. The EW laser is reflected from the prism
surface at an angle of $\theta=45.6^\circ$ ($2^\circ$ above the
critical angle), has a power of $45\,$mW, a $1/e^2$-radius of
540\,$\mu$m and initially a detuning of $\delta_{ew}/2\pi=3\,$GHz with
respect to the D$_2$-line. This leads to a $1/e^2$-decay length of
$\Lambda\approx 500$\,nm and a repulsive optical potential barrier with
a height of $\sim 1$\,mK. This is further reduced to about half of this
value by the attractive van-der-Waals interaction between the atoms and
the dielectric surface \cite{lan96a}.

The hollow beam is generated using an axicon optics \cite{man98} to
create a ring-shaped focus of an inner and outer 1/e-radius of
$r_{HB}=260\,\mu$m and $r_{HB}+\Delta r_{HB}=290\,\mu$m, respectively.
It has a power of 350\,mW and its detuning is in the range between
$-0.3$\,nm and $-2$\,nm. The HB provides a potential barrier on the
order of $100\,\mu$K height.

The weak repumping beam needed for the optical Sisyphus cooling
\cite{soe95} is resonant with the $F=4\rightarrow F'=4$ hyperfine
transition of the $D_2$-line and has an intensity on the order of
1\,$\mu$W/cm$^2$. It is shone on the trapping region from above.

About $2\times10^7$ atoms are loaded into the GOST from a standard
magneto-optical trap in a scheme which includes compression and
precooling to $\sim 10\,\mu$K. Details on loading and other
experimental procedures can be found in \cite{ham00}.

\section{evaporative cooling}\label{secEVAP}
The GOST offers favorable conditions to implement forced evaporative
cooling. In contrast to red-detuned dipole traps used for evaporation
experiments \cite{ada95,eng00}, the spatial compression of the cold
sample essentially results from gravity and is thus not affected when
the optical potentials are ramped down. Moreover, many more atoms are
initially loaded into the GOST as compared to typical red-detuned
traps. Here we describe our first experiments demonstrating the
feasibility of efficient evaporation in the GOST.

The trap is operated at a HB detuning of $-1$\,nm. With an EW detuning
intitially set to a few GHz, Sisyphus cooling provides $N=10^7$\,atoms
at a temperature of $T=10\,\mu$K, and a peak density of
$n_0=6\times10^{11}\,$cm$^{-3}$ \cite{ham00}. For the unpolarized
sample in the seven-fold degenerate $F=3$ ground state this corresponds
to a peak phase-space density of $D=n_0 \lambda_{dB}^3/7\approx10^{-5}$
where $\lambda_{dB}=h/\sqrt{2\pi mk_BT}$ is the thermal de-Broglie
wavelength. Elastic collisions take place at a rate on the order of
$50$\,s$^{-1}$ and, considering the resonant scattering of cesium
\cite{arn97,hop00}, lead to a thermalization time of about 200\,ms.

To implement forced evaporation we lower the EW potential by
ramping up the EW detuning. This simultaneously reduces heating
due to photon scattering and suppresses loss through inelastic
collisions in the presence of blue-detuned light \cite{ham00}.
Within $4.5$ seconds the EW detuning is increased exponentially
from initially 7\,GHz up to 250\,GHz. This is accomplished by
rapid mode-hop free temperature tuning of the EW diode laser. In
the last $2.5$ seconds of the evaporation ramp the intensity of
the hollow beam is reduced from $350\,$mW to $11\,$mW in order to
reduce possible heating by residual light in the dark center of
the hollow beam. The contribution of the HB potential ramp to the
evaporation remains very small.

\begin{figure}[tb]
\begin{center}
\includegraphics[width=7.5cm]{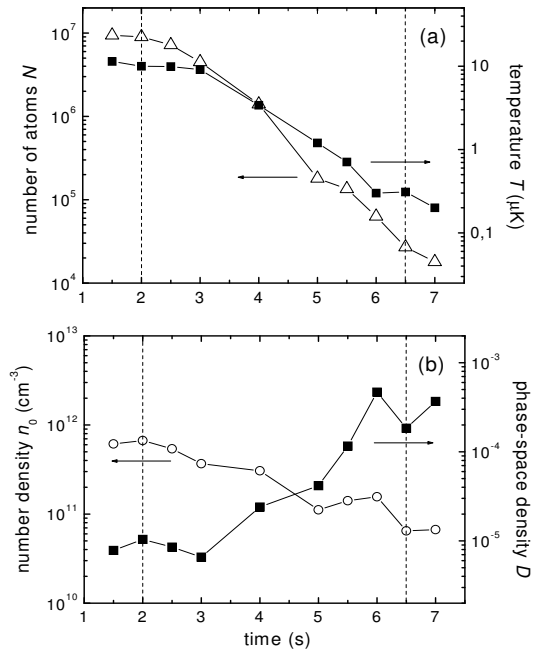}
\end{center}\vspace{-7mm}
\caption{Evaporative cooling in the GOST. The detuning ramp of the EW
starts after two seconds of optical cooling and ends 4.5\,s later,
as indicated by the vertical dashed lines.
The evolution of the number $N$ and the temperature $T$
of the trapped atoms is shown in (a), the corresponding behavior
of peak density $n_0 \propto N/T$ and phase-space density
$D \propto N/T^{5/2}$ is displayed in (b).}
\label{evapAB}
\vspace{3mm}
\end{figure}

The experimental results are shown in figure \ref{evapAB}.
About 1\,s after starting the exponential ramp,
the temperature begins to drop [filled squares in (a)].
At the end of the ramp, it has reached $T \approx 300$\,nK.
This decrease of $T$ by about 1.5 orders of
magnitude is accompanied by a decrease of the particle number $N$
[open triangles in (a)] from
$10^7$ down to $\sim$3$\times10^4$, i.e.\ about
2.5 orders of magnitude.

Although the number density $n_0 \propto N/T$ [open circles in (b)]
decreases by about one order of magnitude, the phase-space density $D
\propto n_0 T^{-3/2} \propto N T^{-5/2}$ [filled squares in (b)] shows
a substantial increase by 1.5 orders of magnitude. At the end of the
ramp, we obtain a phase-space density of $\sim$$3 \times 10^{-4}$.

In the regime of resonant elastic scattering ($T \gtrsim 1\,\mu$K
\cite{hop00}), the relevant cross section scales $\propto T^{-1}$. In
the GOST potential, this leads to a scaling of the elastic scattering
rate and the thermal relaxation rate $\propto N T^{-3/2}$. Therefore
the elastic scattering rate is almost constant for the conditions of
our experiments. However, no runaway regime is reached.

An obvious problem in this evaporation scheme is that, for the applied
exponential ramp, it takes about one second until the EW potential
barrier becomes low enough to start the evaporation. Up to this point
already about 50\% of the particles are lost, presumably by the
collisional mechanism investigated in reference \cite{ham00}. After the
corresponding initial loss of phase-space density the later evaporation
then leads to a gain of almost two orders of magnitude. This already
shows that the potential of evaporative cooling in the GOST is much
larger than we could demonstrate in these first experiments.

\section{heating}\label{secHEAT}
In order to understand the limitations of the evaporation, we have
investigated heating in the GOST. To get quantitative results a
measurement scheme has to be applied in which the average energy per
atom is much smaller than the potential barriers of the trap. That way
one can exclude any effects of evaporation which otherwise would
influence the measurement. Lowering its power by a factor of eight to
$40\,$mW during the first four seconds after the transfer leads to a
strong horizontal energy selection in the sample and thus cooled the
remaining $5\times10^5\,$atoms down to about $3\,\mu$K. To measure
heating without any significant energy selection the HB power is then
switched back to its original value satisfying the condition
$U_{hb}\geq 10 k_B T$.

\begin{figure}[htb]
\begin{center}
\includegraphics[width=7.5cm]{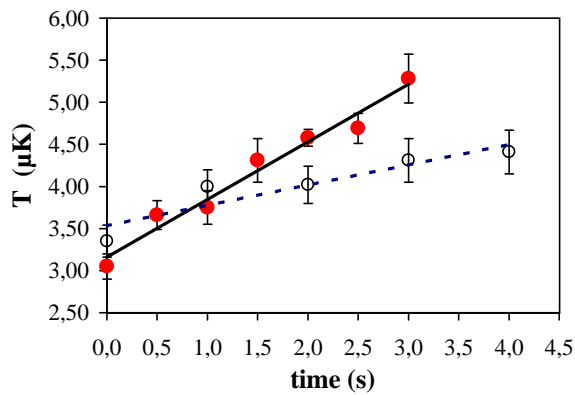}
\end{center}
\caption{Sample temperature versus storage time without ($\bullet$) and
with ($\circ$) a filtering cesium vapour cell in the beam. A linear fit
yields heating rates of $680 \pm 50\,$nK/s and $240 \pm 50\,$nK/s. }
\label{highheat} \vspace{3mm}

\end{figure}

To determine the heating rate, temperature measurements are performed
at various times after the HB power is restored. During this stage the
EW detuning is constantly kept at $\delta_{ew}/2\pi=20\,$GHz as a
compromise to prevent any significant optical cooling effects while
still keeping the EW potential high enough for evaporation not to
occur. Figure \ref{highheat} shows a typical temperature evolution
measurement and the linear fit curve which yields a heating rate of
$\sim 700\,$nK/s.

To calculate the expected heating rate, the mean photon scattering rate
of an atom bouncing on an EW mirror has to be multiplied with the
temperature increase associated with one scattering event. For a
two-level atom the mean scattering rate
is given by \cite{gri00}
\begin{equation}
\Gamma_{sc}=\frac{mg\Lambda\Gamma}{2\hbar\delta_{ew}}\,,
\end{equation}
yielding $\sim$1.2\,s$^{-1}$ under our experimental conditions. In the
geometry of the GOST one scattered photon leads to a heating of
$(2/5)\hbar^2k^2/2m = k_B \times 80\,$nK \cite{gri00}. In combination
with the above scattering rate this would lead to a heating rate of
about 100\,nK/s. However, in the case of cesium atoms with their
hyperfine splitting and the repumping laser applied to keep the atoms
in the $F=3$ state, heating by photon scattering is about twice this
value, so that we can expect a heating rate of $\sim 200\,$nK/s. This
falls short of the experimental result by about a factor of three.


In order to investigate the possibility that a background of {\em
resonant} photons in the EW light causes this discrepancy between
calculation and measurement, we have used a cesium vapor cell heated to
50\,$^{\circ}$C to filter resonant light out of the EW beam. The open
circles in figure \ref{highheat} indicate the resulting temperature
evolution with a slope of 240\,nK/s, which is now indeed consistent
with the expected heating rate. This strongly supports our assumption
that the problem of heating is mainly due to an incoherent background
of resonant photons. In contrast to the case without a filtering cell,
we now also find the residual heating rate to strongly depend on HB and
EW detunings. By tuning the evanescent wave to $40\,$GHz and the hollow
beam to $1\,$nm heating rates as low as $130\,$nK/s were observed.

To further investigate the consequences of photon scattering, we
removed the filtering cesium cell and measured the depolarization of
the ground-state hyperfine population as a function of time. As long as
the repumping beam of the GOST is switched on essentially all atoms
remain in the $F=3$ ground state. Without the repumping beam the sample
gradually depolarizes and atoms are transferred into the $F=4$ state
due to spontaneous Raman scattering.

\begin{figure}[hbt]
\begin{center}
\includegraphics[width=7.5cm,height=5.5cm]{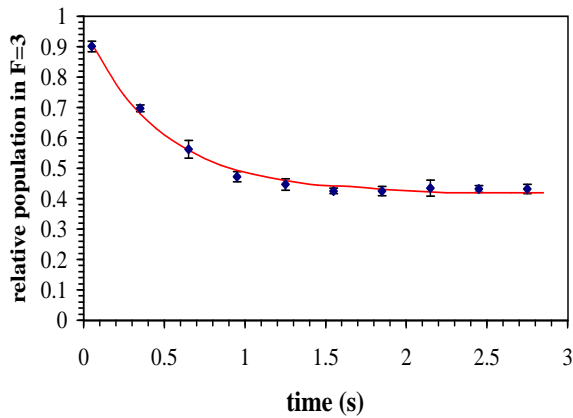}
\end{center}
\caption{Relative population of the F=3 state versus time. From the
decay time of 0.47 seconds and the branching ratios for optical
hyperfine pumping we infer a photon scattering rate of 4.8 photons/s. }
\label{depol} \vspace{3mm}
\end{figure}

To get information on the time constant of the depolarization we
use the same procedure as before to prepare the sample at a
temperature of $3\,\mu$K and then switch off the repumping beam.
After various time intervals a $30\,$ms lasting light pulse of
resonant $F=4\rightarrow F'=5$ light illuminates the sample and
pushes all atoms in the $F=4$ state out of the trap. Measuring the
number of remaining atoms then yields the population of the $F=3$
ground state. Figure \ref{depol} shows the ground state population
normalized to the case without a resonant pulse as a function of
time.

A photon scattering rate of $\sim$5\,s$^{-1}$ can be deduced from the
depolarization rate of the ground-state hyperfine population. With the
repumper present this corresponds to a heating of about $800\,$nK/s.
This again supports our assumption that the observed heating can
essentially be attributed to photon scattering from a very small, but
detrimental fraction of resonant photons present in the EW light.

\begin{figure}[htb]
\begin{center}
\includegraphics[width=7.5cm]{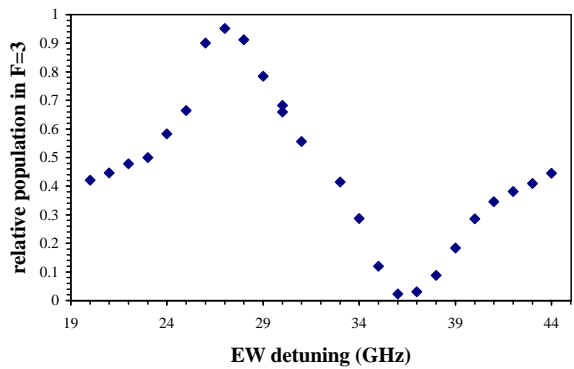}
\end{center}
\caption{Measured population of the $F=3$ ground state after three
seconds depolarization time as a function of the detuning of diode
laser that provides the evanescent-wave light. The surprising
dependence is explained by a amplified spontaneous photons in a
non-lasing side mode of the diode laser leading to hyperfine pumping.}
\label{depolarization} \vspace{3mm}
\end{figure}

To investigate the origin of the resonant photons, we measured the
steady-state ground state population after three seconds of
depolarization time as a function of the EW detuning $\delta_{ew}$.
Figure \ref{depolarization} shows the striking result of this
experiment. The $F=3$ state population exhibits a resonance-like
behavior and takes a value close to 100\% at an EW detuning of
$\delta_{ew}/2\pi=27\,$GHz and vanishes at $\delta_{ew}/2\pi \approx
36\,$GHz. We interpret this as the effect of amplified spontaneous
emission of photons into non-lasing side modes of the diode laser
cavity (mode spacing $\sim 36\,$GHz). As the side mode is tuned in
resonance with the optical transition from the upper state this leads
to an efficient repumping into the $F=3$ ground state and thus to a
relative population of almost one whereas in the opposite case of a
resonance with the optical transition from the $F=3$ state is depleted
by the resonant photons.

The presence of the side modes obviously constitutes a severe problem
for evaporation by ramping up the EW detuning. If a periodic comb of
side modes with a spacing of $36\,$GHz is present during the EW
detuning ramp over 250\,GHz, this leads to several successive
coincidences between side modes and the atomic transition and thus
repeatedly to strong heating. This behavior easily explains the limited
efficiency of evaporation as reported in the preceding section.

\section{Conclusions and outlook}\label{secOUTLOOK}

Our experiments demonstrate evaporative cooling in the gravito-optical
surface trap. A gain in phase-space density of almost a factor of 100
was achieved by ramping up the evanescent-wave detuning. The maximum
phase-space density obtained in these experiments was $\sim$$3 \times
10^{-4}$ at a temperature of about 300\,nK.

As the limiting factor we have identified heating by scattering of
resonant photons contained in the diode-laser light used for the
evanescent wave. Even though the lasing mode of the diode laser is
detuned far enough, amplified spontaneous emission in non-lasing side
modes lead to significant heating. An easy remedy to this problem is
the use of a cesium absorption cell to filter out these photons. First
experiments have indeed shown a substantial reduction of heating with
such a cell.

In other experiments not reported here we have also found strong
indications that at low EW potentials the lifetime of the trapped
sample is severely limited by surface defects of our prism. The
performance of the GOST then strongly depends on the exact position on
the prism surface. Therefore we are now preparing an new set-up
featuring a superpolished high-quality fused-silica prism.

At the lowest temperatures that we have obtained the atoms bounce on
the prism surface with a mean height as low as $\sim$2\,$\mu$m. In this
case, the mean quantum number of the vertical motion is already quite
small (about five to ten). Thus the cold surface gas is not far from
conditions under which the vertical motion will freeze out and the
system will acquire two-dimensional character. With optimized
evaporative cooling, the above improvements, and the further option to
enhance the anisotropic character of the trap by using a second
evanescent wave \cite{ovc91}, a two-dimensional quantum gas in the GOST
with tunable interactions seems to be in experimental reach.

\section*{Acknowledgments}
The reported experiments were performed at the MPI f\"ur Kernphysik in
Heidelberg before our recent move to Innsbruck. We are particularly
indebted to D. Schwalm for continuously supporting our work at the MPI.
We also gratefully acknowledge invaluable support by the Deutsche
Forschungsgemeinschaft in the frame of the Gerhard-Hess-Programm. We
thank D.~Heinzen for his very useful suggestion to use a cesium
filtering cell for the trapping light.

\end{document}